\documentclass[twocolumn,showpacs,preprintnumbers,amsmath,amssymb]{revtex4}
\usepackage{graphicx}
\usepackage[english]{babel}
\usepackage{dcolumn}
\usepackage{bm}

\begin{document}
\title{Proton Threshold  States 
in the $^{22}$Na($p,\gamma$)$^{23}$Mg 
Reaction and Astrophysical Implications}

\author{H. Comisel}\email{comisel@venus.nipne.ro}
\author{C. Hategan}
\affiliation{Institute of Atomic Physics, Bucharest, PO Box MG-6, Romania}

\author{G. Graw} \author{H.H. Wolter}
\affiliation{Department f\"ur Physik, 
Universit\"at M\"unchen, D-85748 Garching, Germany  } 

\date{\today}

\begin{abstract}
Proton threshold states in $^{23}$Mg are important for 
the astrophysically relevant proton 
capture reaction $^{22}$Na($p,\gamma$)$^{23}$Mg.
In the indirect determination of the resonance strength 
of the lowest states, 
which were not accessible by direct methods, some of 
the spin-parity assignments 
remained experimentally uncertain.  
We have investigated these states with Shell Model, 
Coulomb displacement, and Thomas-Ehrman shift calculations.
From the comparison of calculated and observed properties 
we relate  the lowest relevant resonance state at $E_{\mathrm{x}}$=7643 keV 
to an excited $3/2^+$ state 
in accordance with a recent experimental 
determination by Jenkins \textit{et al.}. 
From this we deduce  significantly improved values 
for the $^{22}$Na($p,\gamma$)$^{23}$Mg 
reaction rate at  stellar temperatures below $T_9$=0.1K.
\end{abstract}

\pacs{25.40.Lw, 21.10.Pc, 21.60.Cs, 97.10.Cv}

\maketitle

\section{Introduction}

 In nuclear astrophysics the understanding of  the 
  $rp$ processes as a dominant 
reaction sequence for the Ne-Na cycle is a topic of current 
interest.
Studies of this cycle also have to explain the
anomalous abundance of the $^{22}$Ne  isotope observed 
in the composition of the Orgueil meteorite  
(called Ne-E; E is for extraordinary) 
by Black \textit{et al.} in 1972 \cite{Black72}. 
This isotope is considered
to originate from beta decay  of $^{22}$Na.
 The increased abundance of $^{22}$Ne 
 (with a ratio $^{22}$Ne/$^{20}$Ne $\ge$ 0.67 
 much higher than
the terrestrial $^{22}$Ne/$^{20}$Ne = 0.1)  
points to a scenario of considerable 
production of  $^{22}$Na and relatively weak burning of this
material in the $^{22}$Na($p,\gamma$)$^{23}$Mg reaction 
during the beta decay life time of  $^{22}$Na (T$_{1/2}$=2.6 yr).
Sizeable   $^{23}$Mg production is expected
in the hot Ne-Na cycle, developing in
explosive H-burning locations such as novae. 
The competition between the  production  and the hydrogen burning 
of $^{22}$Na in the proton capture reaction 
 $^{22}$Na($p,\gamma$)$^{23}$Mg 
has been  analyzed in the literature and 
temperatures defining  hot and
cold burning modes have been estimated. 
The  astrophysical aspects of the
$^{22}$Na($p,\gamma$)$^{23}$Mg reaction have been 
outlined in more detail in the work of Seuthe  \textit{et al.}
\cite{Se90} and Schmidt \textit{et al.} \cite{Schmidt95}. 
The results  of the last decade  concerning 
thermonuclear rates for reactions 
 induced by  charged particles are 
systematized in the comprehensive compilation of  Angulo 
\textit{et al.} \cite{Angulo99}.

The astrophysical calculations of the the stellar reaction rate
of the proton capture reaction 
 $^{22}$Na($p,\gamma$)$^{23}$Mg take into account as many as  21 
  resonances in $^{23}$Mg.  
The properties of the lowest three resonances could
not be  determined by direct measurements  
because of the very small capture cross-sections 
at these low proton energies.   
Lower and upper limits of the resonance reaction strengths, however,  
have been  determined in an indirect way,  using the 
proton transfer reaction 
$^{22}$Na($^3$He,$d$)$^{23}$Mg \cite{Schmidt95}.
There the  energy integrated resonance strengths 
have been calculated from the spectroscopic 
factors obtained from the transfer data within the
standard DWBA method, and from the single particle resonance 
widths evaluated from the optical model.

The first resonance, corresponding to the $E_{\mathrm{x}}$=7583 keV state 
in $^{23}$Mg, is very close to the proton capture threshold 
($Q_p$=7579 keV  \cite{Audi95}) and, by barrier penetration arguments,   
its contribution to the reaction rate is negligible.
The strengths of the next two resonances, the
$E_{\mathrm{x}}$=7622 and 7643 keV levels, 
were evaluated to be within the limits 
$5.6 \times 10^{-14} \le\omega\gamma \le
8.8\times 10^{-12}$ meV and 
$1.2\times 10^{-10} \le\omega\gamma \le 3.1 \times 10^{-8}$ meV,
respectively  \cite{Schmidt95}. 
Thus for the reaction rates at
stellar temperatures  below $T_9$ = 0.1
the third resonance 
at $E_{\mathrm{x}}$=7643 keV is important. 
 The uncertainty  of this resonance strength 
 of more  than two orders of magnitude originated 
predominantly from the missing knowledge 
of the value of the transferred  orbital angular momentum.
The ground state of $^{22}$Na has 
spin-parity $3^+$ and the $E_{\mathrm{x}}$=7643 keV state in $^{23}$Mg
has,   according to Endt \cite{Endt98},
spin-parity  $3/2^+$ or $5/2^+$. An assignment of $3/2^+$  
would allow an
orbital angular momentum transfer of $l_{trans}$=2 only, 
whereas an $5/2^+$ assignment allows both 
$l_{trans}$=0 and $l_{trans}$=2.
The transfer angular distribution  of Ref. \cite{Schmidt95} 
had the shape of $l_{trans}$=2, however,
does not allow to exclude the presence of an additional, small 
$l_{trans}$=0 transfer cross section,
which would add incoherently.

 In a recent paper, \cite{Jenkins04,Jenkins06}, 
Jenkins \textit{et al.} have established, that
the 7643 keV level has spin   $3/2^+$. 
The experiment 
used the heavy ion fusion reaction 
$^{12}$C($^{12}$C,$n$)$^{23}$Mg with a subsequent measurement of the radiative
decay  branches. The decay of the 7643 keV state by a 
measured 7196 keV $\gamma$ ray to the first $5/2^+$
excited state supported arguments
for the $3/2^+$
spin assignment. 
Similar arguments fixed the spin assignment of the second 7622 keV state,
which was also largely uncertain, to $9/2^+$.
It is the purpose of this paper 
to give complementary arguments from a different, namely a shell model, 
approach in favor of a $3/2^+$ assignment for the third state, 
and excluding a $5/2^+$ assignment.

\begin{figure}[!b]
   \includegraphics[height=0.48\textheight]{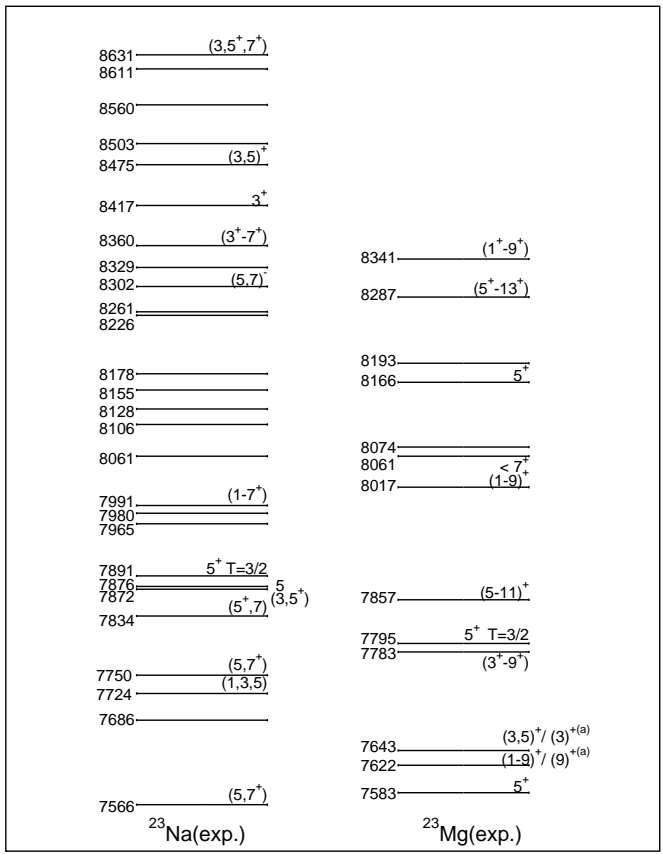} 
    \caption{\label{fig:spectrum} Excitation spectrum in  
    $^{23}$Na and $^{23}$Mg,
   showing experimental data from Ref. \cite{Endt98}, 
       in the range of   astrophysically interesting 
       excitation energies  
    above proton emission  threshold in $^{23}$Mg ($Q_p$=7579 keV). 
    Spins are denoted by 2$J$. $^{(a)}$ Recent spin assignments from Ref.
     \cite{Jenkins04}.}
\end{figure}

In this paper, we apply  the Shell Model (SM)  
to calculate  excitations in 
the  mirror  nuclei  $^{23}$Na and $^{23}$Mg.
We perform calculations of Thomas-Ehrman shifts to relate the levels
of the two mirror nuclei
and we compare the calculated properties of the levels
 with all 
available experimental information in order to 
relate the experimentally observed states with the calculated ones.
The identification of a SM state with the $E_{\mathrm{x}}$=7643 keV state 
in $^{23}$Mg then fixes its spin.

While shell model calculations were applied usually to low lying
 excitations, they have been used 
successfully in the last decade to obtain also
predictions for threshold states of astrophysical 
interest in the  Ne-Na and Mg-Al stellar cycles,
 see \textit{e.g.}  the studies of Champagne \textit{et al.}
\cite{Champagne93} and  Iliadis \textit{et al.} \cite{Iliadis96}.
In the present work, based only on the SM
 and Thomas-Ehrman shift calculations,
we obtain results consistent with  
the results of Jenkins \textit{et al.}, \textit{i.e.} the same spin 
assignment for the 7643 keV level in
 $^{23}$Mg. We consider this as an argument in favor of 
an approach where
shell model and Thomas-Ehrman shift calculations are
extended to energies for high lying proton 
threshold states of astrophysical interest. 
This is the aim of the present work.

Fig.~\ref{fig:spectrum} shows the mirror levels 
in   $^{23}$Na and $^{23}$Mg   
above the proton capture threshold in $^{23}$Mg.
In our approach we start with
SM calculations of $^{23}$Na since, in a fixed basis, these are known 
to  describe better the
levels of the neutron rich than the proton rich, less bound
nucleus \cite{Champagne93}. In addition, 
some levels are better known for  $^{23}$Na.
To compare the mirror levels of $^{23}$Na and $^{23}$Mg nuclei 
we take into account the corresponding
Coulomb and Thomas-Ehrman shifts. 
The theoretical method for the calculation of 
the Coulomb shift of the energy of a level is discussed
in section II. In section III this method is applied to 
 the  isobar analogue mirror states of  $^{23}$Na 
and $^{23}$Mg.
SM assignments for some of the $^{23}$Mg proton capture states 
are obtained and discussed.
In section IV the astrophysical reaction rates are reevaluated and a 
temperature interval, limiting the hot and cold burning
modes of a novae-supernovae scenario, is obtained.

\section{Coulomb and Thomas-Ehrman Shifts in Mirror Nuclei}

The excited states in the mirror nuclei $^{23}$Na and $^{23}$Mg 
are isobaric analog states, the $T_z$=$+1/2$ and 
$T_z$=$-1/2$ members of an isospin $T$=1/2 doublet.  
Thus from the knowledge of 
states of  $^{23}$Na one may identify the  states 
of interest in $^{23}$Mg.
Because of the Coulomb repulsion, 
the latter ones are less bound.
The positive parity states, we are considering here, 
 had  been calculated
with the shell model code OXBASH \cite{Brown85} 
in the $sd$ configuration space. 
These calculations go back to Wildenthal and 
use an established set 
of matrix elements  \cite{Wildenthal84}.
This procedure is appropriate for well bound 
single particle configurations and reproduces  
many features of $sd$-shell nuclei 
in this mass range. 
Because of the fixed basis set of single particle 
wave functions, implied in the
determination of the matrix elements, 
this procedure 
is more appropriate for the study of states in $^{23}$Na 
than for the less bound ones in $^{23}$Mg. 
For the states in $^{23}$Mg we assume 
the same configurations, as obtained for $^{23}$Na. 
To determine their excitation energies we 
calculate Coulomb shifts 
from a  charge-dependent, isospin-nonconserving  
interaction (INC) \cite{Ormand89} 
and in addition 
 the Thomas-Ehrman shifts \cite{TE5152}.
The latter ones take into account, that the 
proton single particle wave 
functions, in particular the $2s1/2$ level, is spatially rather 
extended because of its low binding energy  and 
the absence of a  centrifugal barrier.
Depending on the binding energy and spatial extension
this leads to a reduction of the Coulomb interaction.

More precisely, the Coulomb shifts  have been determined 
applying  the method developed by Herndl \textit{et al.}
 \cite{Herndl95}.
It uses the INC interaction and reproduces well 
 the energy shifts between states 
of predominant $d5/2$ or $d3/2$ configurations.
The Thomas-Ehrman shift 
\cite{TE5152} is derived from calculations with 
wave functions generated in a Woods-Saxon 
single particle potential model \cite{Scherr85,Fortune95}.
Tombrello  \cite{Tombrello66} 
first demonstrated that a one-body potential model
with Coulomb interaction 
can describe the Thomas-Ehrman shift.
For these calculations the shell model wave functions are 
considered as 
a sum of terms, where the valence nucleon is coupled   
to excited states of $^{22}$Na. Then the Thomas-Ehrman shift 
is determined from the  calculated relative shifts of 
$1d5/2$ and $2s1/2$ proton versus neutron 
valence  wave functions at single particle energies, 
 given  by the difference between the energy of the actual state 
and the excitation energy of the core state.
In the  summation the spectroscopic coefficients
determine the weighting factors.

\begin{figure}[t!]
  \centering
    \includegraphics[height=0.22\textheight]{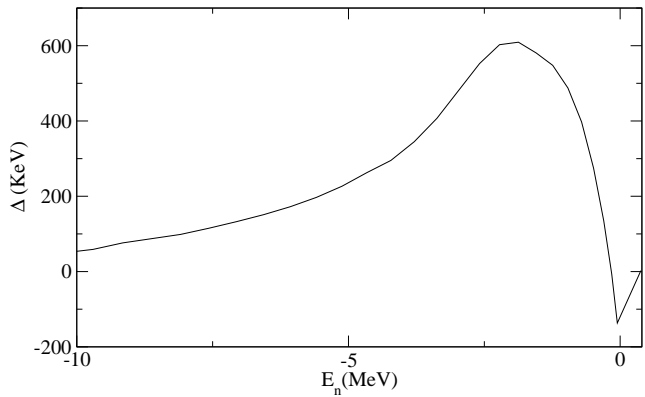} 
    \caption{\label{fig:te_shift} Calculated Thomas-Ehrman shift assuming  pure $2s1/2$ and 
$1d5/2$single particle states for A=23 as a function of the
     $s$-wave neutron energy.   
     }
\end{figure}

Using the Q-values from Ref. \cite{Audi95}, 
the well depth of the Woods-Saxon potential
is chosen to reproduce the single particle energies 
in the neutron rich isotope, $^{23}$Na. 
To determine the energy of the $^{23}$Mg mirror 
levels the same potential is used adding an 
extra Coulomb field of a uniform
spherical charge of radius $r_0A^{1/3}$.
To take into account single particle states 
built onto excited cores,
$s-$wave neutron and proton single particle orbits are 
considered for each of the
states of $^{22}$Na core.
The relative Thomas-Ehrman shift with respect to a 
pure $1d5/2$ state is then 
determined as, 
$\Delta \epsilon_{TE}=\Delta \epsilon_p -\Delta \epsilon_n$,
where $\Delta \epsilon_{n(p)}$ are the single 
particle energy differences between 
$1d5/2$ and  $2s1/2$ neutron (proton) single particle 
states.  
We have used a central Woods-Saxon potential with conventional 
values of the
radius ($r_0$=1.25 fm) and diffuseness (0.65 fm) 
and a uniform charge distribution
($r_C$=1.25 fm).

The Thomas-Ehrman single particle shifts 
$\Delta \epsilon_{TE}$ for the mirror levels 
of $^{23}$Na and $^{23}$Mg
nuclei are a function of the $2s1/2$ 
 neutron single  particle energy. This dependence is shown in 
 Fig.~\ref{fig:te_shift}.
The Thomas-Ehrman shift falls to zero for
higher excitation energies when the $s$-wave neutrons 
reach the threshold. A similar dependence 
has been  obtained for A=13 mass region by Barker \cite{Barker96}.

The energy range we have  considered in the calculation
of the TE shifts, shown in Fig.~\ref{fig:te_shift},   is applicable for  
states  of the  $ ^{22}$Na core up to the
($E_{\mathrm{x}}$=4360 keV, $J^{\pi}=2^+$) level.
As an example, the  level  (7750, ($5/2^+$,$7/2^+$)) 
in $^{23}$Na includes configurations of 
 a $s$-wave neutron coupled to the 
ground state (0., 3$^+$) 
 as well as to the following  excited states in $^{22}$Na: 
 (1951, 2$^+$(\textit{T}=1)),
(1983, 3$^+$), (2968, 3$^+$),(3590, 2$^+$) and (4360, 2$^+$).
The excitation energy, spin and parity for a nuclear level are 
denoted here and in
what follows by ($E_{\textrm{x}}$(keV), $J^{\pi}$). 
The $s$-wave neutron single particle energies 
determined from the Q-values and the 
excitation energies of the core are in the 
($-$4670, $-$310) keV range. 
This interval covers the significant contributions  to
the Thomas-Ehrman shift, see Fig.~\ref{fig:te_shift}.

The resulting relative single particle shifts are 
multiplied by the corresponding
$2s1/2$ spectroscopic coefficients from the SM 
calculation. This term is then 
summed with the SM Coulomb shift using the INC interaction, 
yielding  the total energy displacement.
Finally, this quantity will be added to an experimental 
level in $^{23}$Na with known 
spin-parity assignment, to obtain the predicted 
isobar analogue state in 
the proton-rich nucleus $^{23}$Mg.

\section{Shell Model Calculations and Observed Excited States}

For the mirror nuclei $^{23}$Na and $^{23}$Mg 
we have reliable  assignments for the  
levels  up to  5778 keV in $^{23}$Na and 5711 keV in $^{23}$Mg 
(\textit{e.g.} Table 23j from Ref. \cite{Endt98}).
In Table~\ref{tab:table1} we show the quality of the Coulomb shift calculations of
these low lying bound levels, from which we also obtain an estimate of 
the accuracy of the predictions. The first columns list the low
lying states in $^{23}$Na
and their calculated and experimental excitation energies. 
In the following columns we 
relate this to the experimental energies and predicted values 
$E_{\mathrm{x}}^{TES}$ for the mirror 
nucleus $^{23}$Mg. 
The last column gives the difference of these two quantities. 
The observed mean deviation is about $\pm$60 keV, 
the largest observed one  
is 138 keV. From this we assume, that in the region of 
the proton capture  
states in $^{23}$Mg the calculated energies of the Coulomb shifted 
SM states and 
the experimentally observed ones should have 
twice this mean deviation, that is $\pm$120  keV,
see \textit{e.g.} the
methods used in Ref. \cite{Champagne93} or Ref. \cite{Iliadis96}.
 
\begin{table} [h!]
\caption{\label{tab:table1}Experimental and shell model positive parity states in 
$^{23}$Na  and $^{23}$Mg below the proton threshold. 
For $^{23}$Na we give the experimental energy $E_{\mathrm{x}}$(keV) 
and spin assignment $2J^{\pi}$
and the corresponding shell model states $2J^{\pi}_{SM}$
and the energies 
including the INC interaction $E_{\mathrm{x}}^{INC}$. For $^{23}$Mg
this is put in correspondence with the experimental analogue
states and energies. We further show the energies predicted from 
the Coulomb shift calculations $E_{\mathrm{x}}^{TES}$, which 
are obtained from the relation: 
$E_{\mathrm{x}}^{TES}= E_{\mathrm{x}}(\mathrm{^{23}Na})+
\Delta E_C- \Delta E_{TE}$, 
where $\Delta E_C=E_{\mathrm{x}}^{INC}(\mathrm{^{23}Mg})-E_{\mathrm{x}}^{INC}
(\mathrm{^{23}Na})$ is the pure Coulomb shift 
and $\Delta E_{TE}= \sum{ C^2S(2s1/2)\Delta\epsilon_{TE}} $
is the Thomas-Ehrman shift (see section II). The last column 
gives the difference between the  experimental and predicted
energies for $^{23}$Mg.}
\begin{ruledtabular}
\begin{tabular}{*{7}{c}r}
 & \multicolumn{3}{c}{$^{23}$Na}&\multicolumn{4}{c}{$^{23}$Mg}\\
$2J^{\pi}_{SM}$&$2J^{\pi}$\footnote{Experimental data from Ref. \cite{Endt98}.}
&$E_{\mathrm{x}}^{INC}$&${E_{\mathrm{x}}}$\footnotemark[1] &$2J^{\pi}$\footnotemark[1] &
$E_{\mathrm{x}}^{TES}$& ${E_{\mathrm{x}}}$\footnotemark[1] & $\Delta E_{\mathrm{x}}$\footnote{$\Delta
E_{\mathrm{x}}$(keV)=$E_{\mathrm{x}}(\mathrm{^{23}Mg})-E_{\mathrm{x}}^{TES}$}\\
\hline
$3^+_1 $ &$3^+$ &0    & 0    & $3^+$     & 0	& 0	&\\
$5^+_1 $ &$5^+$ & 411&  440& $5^+$     &  417&  451 & +34\\
$7^+_1 $ &$7^+$ &2119& 2076& $7^+$     & 1972& 2051 & +79\\
$1^+_1 $ &$1^+$ &2297& 2391& $1^+$     & 2297& 2359 & +61\\
$9^+_1 $ &$9^+$ &2785& 2704& $9^+(5^+)$& 2633& 2715 & +81\\
$3^+_2 $ &$3^+$ &2730& 2982& $(3,5)^+$ & 2917& 2908 & -8\\
$5^+_2 $ &$5^+$ &3853& 3914& $(3,5)^+$ & 3726& 3864 & +138\\
$1^+_2 $ &$1^+$ &4289& 4430& $1^+    $ & 4397& 4354 & -43\\
$7^+_2 $ &$7^+$ &4615& 4775& $(1-9)^+$ & 4695& 4685 & -10\\
$5^+_3 $ &$5^+$ &5221& 5379& $(3,5)^+$ & 5327& 5287 & -40\\
$11^+_1$&$11^+$ &5365& 5534&  $\ge 3^+$& 5420& 5456 & +36\\
$5^+_4 $ &$5^+$ &5529& 5742& $5^+    $ & 5713& 5656 & -57\\
$3^+_3 $ &$3^+$ &5724& 5766& $(1-9)^+$ & 5694& 5691 & -3\\
$9^+_2 $ &      &5948& 5778& $(1-9)^+$ & 5626& 5711 & +85\\
\end{tabular}
\end{ruledtabular}
\end{table}

As discussed in the introduction the states of interest for the
astrophysical question of the Ne-Na cycle are the above threshold 
resonances in $^{23}$Mg, in particular the second and third state, whose
spin assignments were uncertain before the work of Ref. \cite{Jenkins04,
Jenkins06}. The aim is to make a correspondence between these 
states and the shell model levels. In Table~\ref{tab:table2} 
we list for the first three states above threshold in $^{23}$Mg
those SM states, whose spin assignments and Coulomb shifted energies fall
into the experimental range of spins and the $\pm$120 keV energy 
interval of the experimental energy. 
These states then are possible candidates to be assigned to the experimental 
proton resonances of $^{23}$Mg. It is seen that this choice is 
rather wide for the second state at 7622 keV and still contains four 
candidates for the 7643 keV state, which is of main interest here.
The determination of the spins of these two levels by Ref. \cite{Jenkins04},
of course, limits these numbers considerably.  
We also list the resonance strength $\omega\gamma$ of these SM states.
We also note, that the only  known \textit{T}=3/2 isobar analogue state  
above proton threshold, 
the $J^{\pi}$=5/2$^{+}$, 7795 keV state, is identified 
reasonably well with the $5/2_9^+$ SM state at 7760 keV 
(not shown in Table~\ref{tab:table2}). 
 
\begin{table} 
\caption{\label{tab:table2} Possible shell model assignations for the first three threshold states in
 $^{23}$Mg. Their  resonance strengths 
were calculated (see section IV) by using SM spectroscopic factors.}
\begin{ruledtabular}
\begin{tabular}{*{2}{c}ll}
$E_{\mathrm{x}}$\footnote{Experimental data from Ref. \cite{Endt98}.}  & $2J^{\pi}$\footnotemark[1] & $ 2J^{+}_{SM}$
& $\omega\gamma $ (meV) \\
\hline
 7583    & $5^+$          & $5_8$ \ $ 5_{10}$               &  $ 2.3 \times 10^{-71}$ \ $  4.7 \times 10^{-71}$    \\
 7622    & $(1-9)^+$      & $1_4$ \ $1_5   $  & $  1.6\times 10^{-16}$ \ $  4.9\times 10^{-15}$  \\ 
         &                & $3_6$ \ $3_7   $  & $  1.0\times 10^{-15}$ \ $  5.6\times 10^{-15}$  \\
         &                & $5_8$ \ $5_{10}$  & $  1.1\times 10^{-12}$ \ $  2.5\times 10^{-12}$  \\
	 &                & $7_6$ \ $ 7_7  $  & $  7.2\times 10^{-12}$ \ $  5.4\times 10^{-13}$   \\
	 &                & $9_7$             & $  4.0\times 10^{-14}$   \\
 7643    & $(3,5)^+$      & $3_6$ \ $3_7   $  & $  1.3\times 10^{-11}$ \ $  7.2\times 10^{-11}$  \\ 
         &                & $5_8$ \ $5_{10}$  & $  1.4\times 10^{-8} $ \ $  3.1\times 10^{-8} $ \\
\end{tabular}
\end{ruledtabular}
\end{table}

\begin{table*} 
\caption{\label{tab:table3} The experimental and theoretical values for: GT transition 
strengths B(GT) corresponding to the $^{23}$Mg $\rightarrow ^{23}$Na  $\beta$ decay and
M1 transition strengths B(M1)$\uparrow$ (units of $\mu_N^2$).} 
\begin{ruledtabular}
\begin{tabular}{cllr|cllr}
\multicolumn{2}{c}{States in $^{23}$Mg}&  \multicolumn{2}{c|}{B(GT)} &
\multicolumn{2}{c}{States in $^{23}$Na}& \multicolumn{2}{c}{B(M1)$\uparrow$} \\
${E_{\mathrm{x}}}$\footnote{Experimental data from Ref. \cite{Endt98}.} &
$2J^{\pi}$\footnotemark[1]
& Experimental\footnote{From Ref. \cite{Fujita2002}.} &
Theoretical\footnote{Using W interaction.} 
& $E_{\mathrm{x}}$\footnotemark[1] & $2J^{\pi}$\footnotemark[1]
 & Experimental\footnote{From Ref. \cite{Fujita2002}, \cite{Endt98}.} & Theoretical\footnotemark[3]   \\
\hline
0    &$3^+    $ & $0.340\pm 0.014$ & 0.541	      	   & 440 &$5^+	  $    & $0.554\pm 0.034$   & 0.483  	       \\
451&$5^+    $ & $0.146\pm 0.006$ & 0.409	      	   &2391 &$1^+	  $    & $0.0017\pm 0.0003$ & 0.026	        \\
2359&$1^+    $ & $0.055\pm 0.004$ & 0.199	      	   &2982 &$3^+	  $    & $0.292\pm 0.041$   & 0.304  	        \\
2908&$(3,5)^+ $ & $0.193\pm 0.011$ & 0.574	      	   &3914 &$5^+	  $    & $0.090\pm 0.015$   & 0.065  	        \\
3864&$(3,5)^+ $ & $0.055\pm 0.004$ & 0.146	      	   &4430 &$1^+	  $    & $1.02\pm 0.07$     & 0.877		        \\
4354&$1^+    $ & $0.250\pm 0.013$ & 0.717	      	   &5379 &$5^+	  $    & $0.33\pm 0.12$     & 0.199		        \\
5287&$(3,5)^+ $ & $0.066\pm 0.005$ & 0.186	      	   &5742 &$5^+	  $    & $0.66\pm 0.04$     & 0.327		        \\
8166&$5^+    $ & $0.290\pm 0.015$ & $(5_{10})^e$ \ \ 0.312&5766 &$3^+	  $    & $0.25\pm 0.04$     & 0.238		        \\
     &  	&		   & $(5_{11})^e$ \ \ 0.058&7133 &$(3,5)^+ $  & $0.31\pm 0.07$     & 0.464		        \\
     &          &                  &  		           & 8360&$(3^+-7^+)$ & $0.290\pm 0.13$    &  $(3_7)$
     \footnote{$2J$ (SM) spins accordingly with $sdpn$ model space.} \  \  0.032  \\
     &          &                  &  		           & 	  &	       & 	            &  $(5_{10})$
     \footnotemark[5] \ \ 0.221   \\
     &          &                  &  		           & 	  &	       & 	            &  $(5_{11})$
     \footnotemark[5] \ \ 0.064   \\
     &          &	           &		           & 8830&$1^+     $  & $0.050\pm 0.022$   & 0.067	 			      \\
\end{tabular}
\end{ruledtabular}
\end{table*}

We now want to determine from the SM point of view
whether the third proton capture resonance state at 
$E_{\mathrm{x}}$=7643 keV has ${J^{\pi}}$=3/2$^+$ or $5/2^+$.
As shown in Table~\ref{tab:table2} we have to consider 
the positive parity SM states  
$J^{\pi}$=$3/2^{+}_6, 3/2^{+}_7, 5/2^{+}_8$ and $5/2^{+}_{10}$, since,
as noted above, the $5/2^{+}_9$ SM state 
is already identified with
the $T$=3/2 isobar analogue state  
above proton threshold.  
To do so we want to argue that we can exclude the two other $5/2^+$ states,
because they have to be assigned to two other states. 
The first resonance state at 
$E_{\mathrm{x}}$=7583 keV  has spin $5/2^+$;
thus we identify this state with the $5/2_8$ SM state.
Now it remains to identify a higher excited state as the $5/2_{10}^{+}$.
As seen in Fig.~\ref{fig:spectrum}, at higher experimental excitation energies in $^{23}$Na 
there are two neighboring states with known and restricted 
 spin and parities:
  $(8417, 3/2^+$) and $(8475, (3/2,5/2)^+$). 
 According to the above procedure 
the SM calculation
relates  these levels in $^{23}$Na
to $(8417 ,3/2_7^+$) and    
to $(8475, 3/2_7^+$ or $5/2_{11}^+$), respectively. 
Because of the 
mutual exclusion we assign  
the 8475 keV level  as $5/2_{11}^+$.
Applying the shift procedure, 
the $5/2_{11}^+$ mirror state in $^{23}$Mg is predicted at  
 $E_{\mathrm{x}}$=8302 keV. 
This excitation energy  exceeds the 
energy of the known  $(8166, 5/2^+$) level in $^{23}$Mg
by 136 keV; thus outside our energy interval. 
Thus we have to identify 
the $(8166, 5/2^+$) experimental level  as 
the $5/2_{10}^+$ SM state.  We will come back to this state below.
Thus the 
$J^{\pi}$=$5/2_{8}^+$, $5/2_{9}^+$ and $5/2_{10}^+$ 
SM states have all been assigned. It follows  that the 
$J^{\pi}$=$3/2^+$ state corresponds to the 7643 keV  level.

To further support the  spin assignments in this energy range, 
a comparison between the
theoretical and available experimental data
for spectroscopic factors, Gamow-Teller $\beta$ 
decay strengths, M1 transition probabilities, and 
gamma-branching ratios are discussed in the following. \\

\noindent a){\it Spectroscopic coefficients}.\\
First we compare SM with experimental 
single proton transfer spectroscopic 
factors from Schmidt \textit{et al.} \cite{Schmidt95}.
The calculated SM spectroscopic factor  $(2J+1)C^2S$=0.02 
 for the $3/2_6^+$ state
 is considerably smaller than the 
value of  $(2J+1)C^2S$=0.10
for the $3/2_7^+$ state. 
The experimental value for the  7643 keV 
level is $(2J+1)C^2S$=0.34 ($l$=2) (see Ref. \cite{Schmidt95}, Table 3.)\\

\noindent b) {\it Gamow-Teller $\beta$ decay transitions.}\\
In a recent paper of Fujita \textit{et al.} \cite{Fujita2002} the GT
transition strengths for $^{23}$Mg$\rightarrow ^{23}$Na $\beta$ 
decay has been
determined,  analyzing the $^{23}$Na($^3$He,$t$)$^{23}$Mg reaction. 
The results are shown in Table~\ref{tab:table3} together
with the SM transition
strengths B(GT) computed with the OXBASH code. 
 The SM values are systematically too large, 
reproduce, however,  the trend of the data.
  The experimental B(GT) for the  
  $(8166, 5/2^+$) level in $^{23}$Mg is 0.29$\pm$ 0.015 while
the predictions for the $J^{\pi}$=$5/2_{10}^+$ and $5/2_{11}^+$ 
SM states are 0.312
and 0.058 respectively.
This observation nicely supports the $J^{\pi}$=$5/2^{+}_{10}$ 
SM assignment, discussed above. \\ 

\begin{table} 
\caption{\label{tab:table4}Experimental gamma ray branching ratio ($\%$) to 
the ground state and the first two excited levels for
 the states, $E_{\mathrm{x}}$=8360 keV, $J^{\pi}$=($3/2^+-7/2^+$) 
 and $E_{\mathrm{x}}$=8166 keV, $J^{\pi}$= 5/2$^+$ 
 in $^{23}$Na and $^{23}$Mg, respectively, (energies in keV).
The values corresponding to  another possible mirror state 
in  $^{23}$Na nucleus, $E_{\mathrm{x}}$=8475 keV, $J^{\pi}$=($3/2-5/2)^+$
are also listed.     }
\begin{ruledtabular}
\begin{tabular}{cccc|ccc}
\multicolumn{4}{c|}{States in $^{23}$Na\footnote{ Exp. gamma ray branching
ratios from  Ref. \cite{Endt98}.}} & \multicolumn{3}{c}{States in
$^{23}$Mg\footnote{Exp. gamma ray branching ratios from  Ref. \cite{Se90}.}}\\
$E_{\textrm{x}}$ & $2J^{\pi}$ & 8360   & 8475 & $E_{\textrm{x}}$& $2J^{\pi}$ & 8166\\
\hline
 0.0  &$3^+$ & 53 $\pm$3 &      & 0.0  & $3^+$ & 65 $\pm$5 \\
 440  &$5^+$ & 32 $\pm$3 & (50) & 451  & $5^+$ & 19 $\pm$2 \\
 2076 &$7^+$ & 15 $\pm$2 &      & 2051 & $7^+$ & 16 $\pm$2 \\
other &      &         & (50)   &      &       &           \\
levels&      &         & unknown&      &       &           \\
\end{tabular}
\end{ruledtabular}
\end{table}

\begin{table}
\caption{\label{tab:table5}Experimental and Shell Model  $5/2^+$ states in $^{23}$Mg.
The $E_{\mathrm{x}}^{TES}$ predicted energies  written in brackets were obtained 
from assumed  assignments of the analogue states in $^{23}$Na.}
\begin{ruledtabular}
\begin{tabular}{*{7}{c}r}
$2J^{\pi}_{SM}$&${2J^{\pi}}^a$&$E_{\mathrm{x}}$\footnote{Experimental data from
Ref. \cite{Endt98}.} & 
${E_{\mathrm{x}}}^{TES}$& $\Delta E_{\mathrm{x}}$\footnote{$\Delta E_{\mathrm{x}}$(keV)=$E_{\mathrm{x}}(^{23}Mg)-E_{\mathrm{x}}^{TES}$} \\ 
\hline
$5^+_1    $& $5^+$	& 451  & 417  & +34	\\ 
$5^+_2    $& $(3,5)^+$  & 3864 & 3726 & +138 \\ 
$5^+_3    $& $(3,5)^+$  & 5287 & 5327 & -40  \\ 
$5^+_4    $& $5^+    $  & 5656 & 5713 & -57  \\ 
$5^+_5    $& $5^+$	& 6568 &(6632)& (-64) \\ 
$5^+_6    $& $5^+$	& 6899 &(6856;6906) &  (43;-7) \\ 
$5^+_7    $& $5^+$	& 6984 &(6950)& (+44) \\ 
$5^+_8    $& $5^+$	& 7583 & 7466 to 7702&  \\ 
$5^+_9    $& $5^+$,T=3/2& 7795 &  7760& +35 \\ 
$5^+_{10} $& $5^+$	& 8166 &  8100& +66\\ 
$5^+_{11} $&  &8193 to 8420  &  8302      &  \\ 
\end{tabular}
\end{ruledtabular}
\end{table}

\begin{table*} 
\caption {\label{tab:table6} The resonance strengths for the three lowest states in $^{23}$Mg above the proton threshold. 
}
\begin{ruledtabular}
\begin{tabular}{*{6}{c}}
$E_{\textrm{x}}$(keV)\footnote{From Ref. \cite{Endt98}.}  & $2J^{\pi}$\footnotemark[1] &  
\multicolumn{4}{c}{$\omega \gamma (meV)$\footnote{According to the spin assignments and resonance energies used in the cited works.}}\\
\cline{3-6}
  &                       & \multicolumn{2}{c}{Ref. \cite{Schmidt95}} & Ref. \cite{Jenkins06} & Present \\
\cline{3-4}
  &                                                             &      low              &  high   &       &   high \\
\hline
7583    & $5^+$               &   0			 & $1.3 \times 10^{-63}$  &  &  \\
7622    & $(1-9)^+$           &   $ 5.6\times 10^{-14}$ & $8.8\times 10^{-12}$ 
 & $1.7_{-1.1}^{+2.5}\times 10^{-13}$ & $1.0\times 10^{-13}$
 \footnote{According to the spin assignment of Ref. \cite{Jenkins06}.} \\
7643    & $(3-5)^+$           &   $ 1.2\times 10^{-10}$ & $3.1\times 10^{-8}$    &  $2.2_{-1.4}^{+3.0}\times 10^{-9}$ & $2.2\times 10^{-10}$ \\
\end{tabular}
\end{ruledtabular}
\end{table*}
\noindent c) {\it M1 gamma transitions.}\\
The M1 transition strengths B(M1)$\uparrow$ from the 
ground to the excited
states in $^{23}$Na are  calculated from the 
SM transition densities and
are  compared to the experimental ones of Fujita \textit{et al.} 
\cite{Fujita2002}, and are also shown in Table~\ref{tab:table3}.
The theoretical and experimental values generally are in good agreement.
 In particular, the experimental value of
(B(M1)=0.290$\pm$0.13) for the $E_{\mathrm{x}}$=8360 keV in $^{23}$Na
agrees well with 
(B(M1)=0.221) for the  $5^{+}_{10}$ SM state. 
This then supports the  $5/2^{+}_{10}$
SM assignment for the 
$E_{\mathrm{x}}$=8166 keV excited state in $^{23}$Mg, as discussed. 
The predicted excitation energy of the mirror 
state  in $^{23}$Mg exceeds the experimental one 
 by  $\Delta E_{\mathrm{x}} \simeq70 $ keV  only.\\

\noindent d) {\it Gamma-branching ratios.}\\
  The mirror relation of the $E_{\mathrm{x}}$=8360 keV state in 
 $^{23}$Na and the $E_{\mathrm{x}}$=8166 keV state in $^{23}$Mg is 
 further supported by the  experimental gamma branching
ratios to low lying state in these nuclei. This is shown
in Table~\ref{tab:table4} together with data from the literature.
The agreement is surprisingly good.  
In contrast, another candidate  analogue state in $^{23}$Na at 
$E_{\mathrm{x}}$=8475 keV 
 has appreciable deviations. 

Summarizing  the arguments (a)-(d), the present analysis  
further supports the $5^{+}_{10}$ SM assignment to the 
$E_{\mathrm{x}}$=8166 keV 
state in $^{23}$Mg. This in turn supports the 
 $J^{\pi}$=3/2$^+$ assignment of the 
7643 keV  level. We thus see  that purely from SM arguments 
we are able to assign a spin to the 7643 keV level in $^{23}$Mg.
At similar argument would be more difficult with the second 
7622 keV level, because of many SM possibilities listed in 
Table~\ref{tab:table2}.

From  the above arguments we are able to assign the first ten
$5/2^+$ states in the SM calculations to definite states in 
$^{23}$Mg. This is shown in Table~\ref{tab:table5} for all the states up to 
about 8 MeV excitation energy. One can 
also make some spin assignments more definite. 
The second  and third states of $^{23}$Mg at
$E_{\mathrm{x}}$=3864 and 5287 keV 
are IAR of the states at $E_{\mathrm{x}}$=3914 and 5379 keV in $^{23}$Na
\cite{Endt98} with spins $5/2^+$. Consequently  these
 two states of $^{23}$Mg are also $5/2^+$ states. All other higher
 states from Table~\ref{tab:table5} have a well-assigned $5/2^+$ spin; 
there is no place for assignment of $5/2^+$ spin to the 7643
level in $^{23}$Mg, lending further support to the assignment of $3/2^+$.
In addition this table demonstrates that SM calculations can be useful to 
identify experimental levels also at excitation energies above the
threshold, and can thus be of use in astrophysical considerations.

\section{Resonance Strengths and Stellar Reaction Rates}

The resonance strengths for the states just above the
 proton threshold has been determined by the relation,
\begin{eqnarray}
\omega\gamma=\frac{2J+1}{2(2J_0+1)}\frac{\Gamma_p\Gamma_{\gamma}}{\Gamma_{tot}}
\approx \frac{2J+1}{2(2J_0+1)}\Gamma_p  \nonumber
\end{eqnarray}
where $J_0$ is  the spin of $^{22}$Na target, 
$J$ is the spin of the resonant state, 
while $ \Gamma_p$, $\Gamma_{\gamma}$ and $\Gamma_{tot}$ are the
partial proton width in the entrance channel, 
the partial $\gamma$ width in the 
exit channel and the total width, respectively. 
The last equality is valid for $\Gamma_{\gamma}\ll\Gamma_p$, as here.
The proton widths are determined as product of 
the spectroscopic factors (from SM 
calculations or from experimental data)
and the single particle width 
$\Gamma_p=C^2S\times\Gamma_{sp}$,
 where $C$ is the isospin Clebsch-Gordan coefficient. 
The single particle width is derived within the optical model as a product
of the penetration factor and the Wigner unit $\Gamma_{sp}=P |u(r_a)|^2$
\cite{dwuck4}.
The same  parameters of the optical potential have been used, as for
the Thomas-Ehrman shift evaluation.

The experimental information on the lowest three states in 
$^{23}$Mg above the proton threshold which have been 
included in the calculation of the 
reaction rate is shown in Table~\ref{tab:table6}.
The excitation energies and the knowledge about the
spin-parities  $J^{\pi}$  are given in columns 1 and 2, 
accordingly with Ref. \cite{Endt98}.
The remaining columns give the
experimental information on the resonance strength. 
In column 3 and 4 we give the low 
and high limits of Ref. \cite{Schmidt95} (which includes the uncertainty 
due to the uncertainty of the spin of the second and third states), 
 and in column 5 the values quoted by Jenkins \textit{et al.} in their last publication
\cite{Jenkins06}. In the last column the results for the higher limits of the present work 
are given. The lower limits are nearly the same as the lower ones from Ref. \cite{Schmidt95}
and we did not list them again in  Table~\ref{tab:table6}.
The low and high limits of the resonance 
strengths from Ref. \cite{Schmidt95} for the third level were mainly due to
 the possible \textit{l}=2 and \textit{l}=0  values of the transferred orbital momenta. 
The elimination 
 of \textit{l}=0 orbital now reduces strongly the range of  possible values of the
  resonance strength.
Using the experimental spectroscopic coefficient for \textit{l}=2 and the prescription
from Ref. \cite{Schmidt95}  for  the upper limit of $\omega\gamma$, 
the maximum value of the resonant strength is now 
$\omega\gamma(expt.)_{high}=2.2\times 10^{-10}$ meV. 

Of particular interest is the strength of the third state. While the limits 
for the strength in Refs. \cite{Schmidt95, Angulo99} are almost the same and
 rather wide, 
these are quite narrowly defined in the present work. The values quoted in
Refs. \cite{Jenkins04} and \cite{Jenkins06} are different. In the table and in
the ensuing calculations we 
have used the latest values which are considerably higher that those of the
present work, even though now the same spin assignment is used.

The astrophysical reaction rates have been computed according to 
the formalism of narrow resonances 
\cite{Rolfs88}, where
the contributions are 
calculated separately for each of the  analyzed levels.
The resonant reaction rates depend on the  
resonance energies $E_r$ and the  
resonance strengths $\omega\gamma$,
as well as on the $T_9$  temperature,  \cite{Fowler75},
\begin{eqnarray}
N_A<\sigma v>_r= 1.54\times 10^{11}(A T_9)^{-3/2} (\omega\gamma) 
exp \left( -\frac{11.605E_r}{T_9} 
\right)  \nonumber
\end{eqnarray}
Here, A is the reduced mass, $A=A_pA_T/(A_p+A_T)$, 
$A_p$ is the projectile mass, and $A_T$ 
is the
target mass. The reaction rate
 $N_A<\sigma v>_r$ is expressed in units of 
 cm$^3$s$^{-1}$mol$^{-1}$ if the
strengths are given in meV, the resonance energies in MeV, and 
temperatures $T_9$ in 10$^9 K$.

\begin{figure}[t!]
  \centering
    \includegraphics[height=0.23\textheight,]{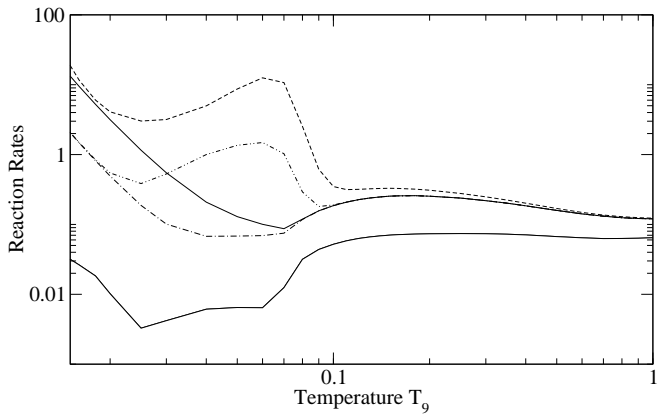}
    \caption{\label{fig:rrate}  Upper and lower limits of the reaction rate 
     versus stellar temperature $T_9$ for the
    $^{22}$Na($p,\gamma$)$^{23}$Mg reaction.  Using the adopted values 
    of Ref. \cite{Angulo99} the lower solid and the upper dashed 
    curves are obtained.  We also show the modifications of the upper limit
    of the reaction rate due other assumptions. 
    The upper solid line is the reaction rate of this work taking into account 
    the $3/2^+$ spin assignment of the third level. The dash-dotted curve is 
    obtained from this solid curve by using for the second state the
    spin assignation given by Ref. \cite{Jenkins06}. The dash-double-dotted curve uses
    in addition the strength of the third state as given in Ref. \cite{Jenkins06}.
    The reaction rates are normalized to 
    those of Caughlan and Fowler, Ref. \cite{Ca88}.
     }
\end{figure}

In Fig.~\ref{fig:rrate} the upper and lower limits of the reaction rate 
with the various assumptions are shown,
normalized to the ones calculated by Caughlan and Fowler, 
Ref. \cite{Ca88}.
The lower solid line and the highest 
dashed lines correspond to the reaction rate obtained with 
the lower and upper values, respectively, for the strength functions
adopted by Ref. \cite{Angulo99}. 
The other curves represent the upper limits with the other assumptions.
The upper solid line is the reaction rate of this work taking into account 
the spin $3/2^+$ assignment of the third level. The dash-dotted curve is 
obtained from this solid curve by using for the second state
 the spin assignment of Ref. \cite{Jenkins06} and corresponding spectroscopic 
factor of Ref. \cite{Schmidt95}.
 The dash-double-dotted curve uses
in addition the strength of the third state as given in Ref. \cite{Jenkins06}.
Two aspect will be noted:  first, the upper limit of Jenkins 
\textit{et al.}, is much 
higher, due to the higher resonance strength. Secondly,
the more precise limits for the second state 
do not much affect the reaction rate in the astrophysically interesting 
region above $T_9 > 0.05$.

\begin{figure}[!t]
  \centering
   \includegraphics[height=0.23\textheight,angle=0]{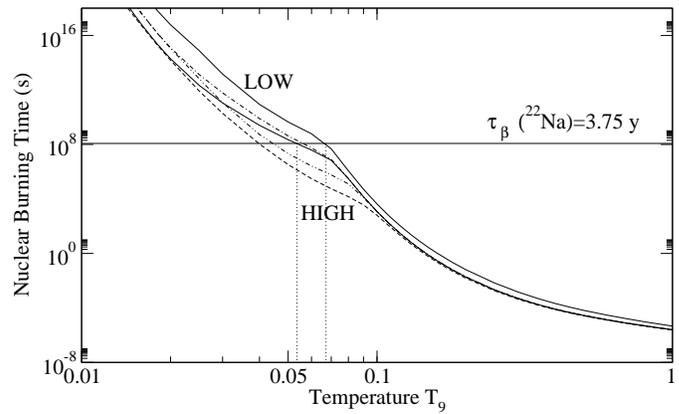}
    \caption{\label{fig:ltime} The $\beta$-decay lifetime $\tau_{\beta}$($^{22}$Na) 
    and the limits of lifetime of $^{22}$Na 
    against proton capture, $\tau_p$($^{22}$Na),
     versus temperature, calculated for a pure 
     hydrogen composition
    and a density of $\rho$=1000 g/cm$^3$. The curves were obtained
     from the reaction rates shown in Fig. 3.
}
\end{figure}

From the above limits of the reaction rates,  the limits of the 
lifetime  against
proton capture  $\tau_p(^{22}$Na)=$(\rho X_H$ $N_A<\sigma v>)^{-1}$
versus stellar temperature $T_9$,
have been calculated assuming a stellar density of $\rho$=1000g/cm$^3$ and a hydrogen
 mass fraction 
$X_H$=1. These limits are compared to the $\beta$-decay lifetime $\tau_{\beta}$($^{22}$Na) 
in Fig.~\ref{fig:ltime}. The intersections with this line determine the limits of temperature 
corresponding to the cycle switches between cold 
and hot burning modes. 
The lines represent the different assumptions on the burning rates in
Fig.~\ref{fig:rrate}, 
where the same line signatures are used as there, except, 
of course, that upper and lower limits are now interchanged.
For the reaction rates of Ref. \cite{Angulo99} 
(solid upper and dashed lower line) the limits 
$T_9$=0.039 and  $T_9$=0.068 had  already been calculated in Ref. 
\cite{Schmidt95}. 
Using the
 upper limit for the reaction rate of this work (solid lower line),
  the temperature interval defining the burning mode lies at
$T_9$=0.055 to $T_9$=0.068 (marked by dotted vertical lines in
 Fig.~\ref{fig:ltime}), thus considerably sharpening the lower limit. 
   Using instead the smaller values corresponding to the $9/2^+$ spin assignment
 of Ref. \cite{Jenkins06} for
 the second resonance (dash-dotted line) does not significantly
  change the lower limit of the cycle switching temperature. Finally using
 the value of Ref. \cite{Jenkins06}  for the third resonance strength
 (dash-double-dotted line) significantly lowers the lower limit.

\section{Conclusions}

In view of the astrophysical importance of the $^{22}$Na($p,\gamma$)$^{23}$Mg
reaction, 
we have analyzed the spin and parity of
the lowest levels above the proton capture threshold in the  
$^{23}$Mg nucleus with the help of Shell Model calculations. 
The predicted SM levels for the mirror nuclei 
$^{23}$Na and $^{23}$Mg 
were calculated using the  Wildenthal interaction 
in the appropriate $sd$ space model.
The isobar analogue nuclear levels were matched 
theoretically by calculating the Coulomb displacements,
including the Thomas-Ehrman shifts.

From the comparison of the 
SM and experimental 
energy levels and other spectroscopic data, in particular 
spectroscopic coefficients, Gamow Teller beta 
decay and B(M1) gamma transition amplitudes 
and experimental gamma-branching ratios, 
we deduced that the third state
just above the proton threshold in $^{23}$Mg 
at $E_{\mathrm{x}}$=7643 keV has a 3/2$^+$ spin assignment. 
This is in agreement with dedicated experimental studies by Jenkins \textit{et al.} 
\cite{Jenkins04,Jenkins06}, 
who also argued for a 3/2$^+$ value. Their 9/2$^+$ spin assignation 
for the lower $E_{\mathrm{x}}$=7622 keV level, the second above threshold, cannot be
uniquely predicted by the present SM approach. 
However it does not significantly contribute to the astrophysical reaction rate
for $T_9>$0.05 or for the expected lower limit 
of the cycle switching temperature. 
Using the lower limit of our resonance strength 
for the third, \textit{i.e.} 3/2$^+$,  resonance above the proton 
threshold, we obtain  significantly reduced upper  
limits for the thermonuclear 
reaction rate below $T_9$=0.1. 
Consequently,  the temperature interval 
defining the stellar burning modes for a novae-supernovae 
scenario is  accordingly narrowed.

More generally, we have demonstrated the usefulness of SM calculations in obtaining assignments of high lying levels, even above threshold, which might be of interest in astrophysical reaction scenarios.

\begin{acknowledgments} 
C.H. acknowledges  support of the A. von Humboldt--Stiftung and the hospitality at the Ludwig-Maximilians-University of Munich, while working on
final variant of this work. H.C. acknowledges the hospitality of the Max-Planck-Institute for Extraterrestrial Physics (MPE) at Garching.
\end{acknowledgments}


\begin{thebibliography}{99}
\bibitem{Black72} D.C. Black, Geoch. Cosmoch. Acta {\bf 36}, 347 (1972).
\bibitem{Se90} S. Seuthe, C. Rolfs, U. Schr$\mathrm{\ddot{o}}$der, W.H. Schulte, E. Somorjai,
H.P. Trautvetter, F.B. Waanders, R.W. Kavanagh, H. Ravn, M. Arnould, G. Paulus, Nucl. Phys.
{\bf A514}, 471 (1990).
\bibitem{Schmidt95} S. Schmidt, C. Rolfs, W.H. Schulte, H.P. Trautvetter,
R.W. Kavanagh, C. Hategan, S. Faber, B.D. Valnion, G. Graw, Nucl. Phys. 
{\bf A591}, 227 (1995).
\bibitem{Angulo99} C. Angulo, M. Arnould, M. Rayet, P. Descouvemont, D. Baye,
C. Leclerq-Willain, A. Coc, S. Barhoumi, P. Aguer, C. Rolfs, R. Kunz, J.W. Hammer, A. Mayer, 
T. Paradellis, S. Kossionides, C. Chronidou, K. Spyrou, S. Degl'Innocenti, G. Fiorentini, 
B. Ricci, S. Zavatarelli, C. Providencia, H. Wolters, J. Soares, C. Grama, J. Rahighi, 
A. Shotter, M. Lamehi Rachti, Nucl. Phys. {\bf A656}, 3 (1999).
\bibitem{Audi95} G. Audi, A.H. Wapstra, Nucl. Phys. {\bf A595}, 409 (1995).
\bibitem{Endt98} P.M. Endt, Nucl. Phys. {\bf A633}, 1 (1998); Nucl. Phys. {\bf A521}, 1 (1990).
\bibitem{Champagne93} A.E. Champagne, B.A. Brown, R. Scherr, Nucl. Phys. {\bf A556}, 123 (1993).
\bibitem{Iliadis96} C. Iliadis, L. Buchmann, P.M. Endt, H. Herndl, M. Wiescher, Phys. Rev.
 C {\bf 53}, 475 (1996).
\bibitem{Jenkins04} D.G. Jenkins, C.J. Lister, R.V.F. Janssens, T.L. Khoo,
E.F. Moore, K.E. Rehm, B. Truett, A.H. Wuosmaa, M. Freer,
B.R. Fulton, J. Jose, Phys. Rev. Lett. {\bf 92}, 031101 (2004).
\bibitem{Jenkins06} D.G. Jenkins, C.J. Lister, R.V.F. Janssens, T.L. Khoo,
E.F. Moore, K.E. Rehm, D. Seweryniak, A.H. Wuosmaa, T. Davinson, 
P.J. Woods, A. Jokinen, H. Pentilla, G. Martinez-Pinedo, J. Jose, 
Eur. Phys. J. A {\bf 27}, 117-121 (2006); \\
D.G. Jenkins, B.R. Fulton, C.J. Lister, R.V.F. Janssens, T.L. Khoo,
E.F. Moore, K.E. Rehm, B. Truett, A.H. Wuosmaa, M. Freer, J. Jose,
Nucl. Phys. {\bf A758}, 749c-752c (2005).

\bibitem{Brown85} B.A. Brown, A. Etchegoyen and W.D.M. Rae, The 
computer code OXBASH, MSU-NSCL Report No. 524.
\bibitem{Wildenthal84} B.H. Wildenthal, Prog. Part. Nucl. Phys. {\bf 11}, 5 (1984).
\bibitem{Ormand89} W.E. Ormand, B.A. Brown, Nucl. Phys.
{\bf A491}, 1 (1989); W.E. Ormand, Phys. Rev. C {\bf 55}, 2407 (1997).
\bibitem{TE5152} J.B. Ehrman, Phys. Rev. {\bf 81}, 412 (1951); 
R.G. Thomas, Phys. Rev. {\bf 88}, 1109 (1952), {\bf 81}, 148 (1951).
\bibitem{Herndl95} H. Herndl, J. G$\mathrm{\ddot{o}}$rres, M. Wiescher,
B.A. Brown, L. Van Wormer, Phys. Rev. C {\bf 52}, 1078 (1995).
\bibitem{Scherr85} R. Sherr and G. Bertsch, Phys. Rev. C {\bf 32}, 1809 (1985).
\bibitem {Fortune95} H.T. Fortune, Phys. Rev C {\bf 52}, 2261 (1995).
\bibitem{Tombrello66} T.A. Tombrello, Phys. Lett. {\bf 23}, 134 (1966).
\bibitem{Barker96} F.C. Barker, Phys. Rev. C {\bf 53}, 2539 (1996).
\bibitem {dwuck4} P.D. Kunz, computer code DWUCK 4, unpublished.
\bibitem {Fujita2002} Y. Fujita, Y. Shimbara, I. Hamamoto, T. Adachi, G.P.A. Berg, H. Fujimura,
H. Fujita, J. G$\mathrm{\ddot{o}}$rres, K. Hara, K. Hatanaka, J. Kamiya, T. Kawabata, Y. Kitamura, 
Y. Shimizu, M. Uchida, H.P. Yoshida, M. Yoshifuku, M. Yosoi, Phys. Rev. C {\bf 66}, 044313 (2002).
\bibitem {Rolfs88} C.E. Rolfs and W.S. Rodney, Cauldrons in the Cosmos (University of
Chicago Press, Chicago, 1988).
\bibitem {Fowler75} W.A. Fowler, G.R. Caughlan and B.A. Zimmermann, Annu. Rev. Astro. Astrophys,
{\bf 13}, 69 (1975).
\bibitem {Ca88} G.R. Caughlan and W.A. Fowler, At. Data Nucl. Data Tables {\bf 40}, 283 (1988).
\end{thebibliography}
\end{document}